\documentclass[twocolumn,showpacs,preprintnumbers,amsmath,amssymb]{revtex4}
\usepackage[latin1]{inputenc}
\usepackage[english]{babel}
\usepackage{graphicx}
\begin{document}
\title{From inflation to late time acceleration with a decaying vacuum coupled to radiation or matter}
\author{Stéphane Fay\footnote{steph.fay@gmail.com}}
\affiliation{
Palais de la Découverte\\
Astronomy Department\\
Avenue Franklin Roosevelt\\
75008 Paris\\
France}
\begin{abstract}
We consider General Relativity with matter and radiation, one of these fluids being coupled to vacuum. We find that Universe dynamics starts by an inflation phase if the coupled fluid has a negative energy density at early time. Then, there is always a finite scale factor singularity but when vacuum and matter are coupled and matter density behaves like a negative radiation density. Moreover, the convergence to the $\Lambda CDM$ model is clearly easier to reach when vacuum is coupled to matter rather than to radiation. Two classes of theories are studied to illustrate these results.
\end{abstract}
\maketitle
\section{Introduction} \label{s0}
Universe expansion is presently accelerating as shown by observations like supernovae surveys \cite{Rie98,Per99}, baryon acoustic oscillations \cite{Eis05} or CMB\cite{Ade13A}. This acceleration took place some few billions years ago and could be due to a dark energy. Before this time, expansion was decelerated, at least to the nucleosynthesis time. This Universe dynamics is well described by the $\Lambda CDM$ model which is the simplest type of dark energy. However, another period of expansion acceleration, named inflation, is also generally assumed at very early time to solve the horizon, flatness and monopole problems\cite{Gut81,Bra01,Lin04}. Although other paradigms have been suggested to solve them, such as a torsion of spacetime\cite{Pop11}, string gas cosmology\cite{Bra89} or matter bounce\cite{Pet08}, inflation is in very good agreement with the most recent measurements of the CMB\cite{Ade13}. It is thus one of the most appealing candidate to describe the very early Universe. Since the $\Lambda CDM$ model cannot reproduce two periods of accelerated expansion, one at early time and the other at late time, it is necessary to go beyond it to find some cosmological models describing the full Universe dynamic from early time inflation to late time acceleration\cite{Lim12,Lim13}.\\
One way to get such a scenario is to assume a coupling between vacuum energy\cite{Car01,Sol13} and some fluids, leading to a decaying vacuum energy\cite{Fre87}. This is what is done in \cite{Lim13} where a vacuum energy density is modelised as a power series of the Hubble function and coupled to both radiation and matter. It allows to get a complete cosmological scenario with a spacetime emerging from an initial de Sitter stage and subsequently evolving into the radiation, matter and dark energy dominated epochs.\\
The idea to consider a coupling between dark energy or inflation fields, generally modeled by some scalar fields or some perfect fluid equations of state\footnote{We will not consider in this paper alternative theories of gravity like $f(R)$ theories or scalar-tensor theories but will stay in the realm of General Relativity.}, with matter or radiation is not new\cite{Ell89}. It can be in agreement with observations\cite{Pou13, Pet13} and it allows to solve important cosmological problems. Hence, the coupling of dark energy with dark matter is used to alleviate the coincidence problem, i.e. the fact that the energy densities of these two fluids have the same amplitude today\cite{Ame99,Zim01,Ame01,Ame06}. As in absence of coupling, it also allows to alleviate the cosmological constant problem\cite{Wei89} when dark energy density decreased with time. From the quantum viewpoint of particle physics, it is natural that the inflaton field interacts with other fields. For instance couplings of the inflaton with scalar and spinor fields naturally arise in gauge theories with spontaneously broken symmetry \cite{Muk05}. Concerning the coupling of inflaton to fluids like matter or radiation, it is able to lead to a graceful exit from inflation with particles production. Among others, this is the case in the warm inflation scenario\cite{Bas09} where the inflaton decayed into radiation. Indeed, particles production during inflation can even be traced back to a pre-inflation paper by L.Z. Fang\cite{Fan80} showing that dissipative process near the Higgs phase transition could allow to produce cosmic entropy. Although it is now believed that the Higgs field is not the inflaton, the idea of an inflaton dacaying into particles to end the inflation epoch and enter into radiation era is an active field of research since many decades\cite{Mos85,Yok88,Ber06}.\\
In the present paper, we consider a decaying vacuum energy as in \cite{Lim13} but coupled only to matter or radiation and without specifying any quantity such as the Hubble function. We assume an expanding Universe and a positive coupling function between vacuum and one of the two fluids. We use dynamical system method\cite{BilCol00} to determine the properties of matter, radiation and dark energy necessary to lead Universe from an inflation epoch to a $\Lambda CDM$ one. We find that when vacuum is coupled to matter (radiation), the matter (respectively radiation) energy density has to be negative at early time to allow an inflation epoch. Moreover the singularity is always a finite scale factor singularity but when matter is coupled to vacuum and its energy density behaves like this of radiation with a negative sign. When vacuum is coupled to radiation, it is difficult to reach a $\Lambda CDM$ epoch early in Universe history. However, this is far more easier when vacuum is coupled to matter. Then, the convergence to the $\Lambda CDM$ model is possible as soon as the nucleosynthesis epoch and even before. We illustrate these results with two simple forms of coupling functions.\\
The plan of the paper is as follows. In section \ref{s1} we consider the coupling of vacuum with radiation. In section \ref{s2} we consider the coupling of vacuum with matter. We discuss about these results and conclude in section \ref{s3}.
\section{Vacuum energy coupled to radiation} \label{s1}
We consider a positive coupling between vacuum and radiation such as vacuum is cast into radiation.
\subsection{Field equation} \label{s11}
When vacuum is coupled to radiation, the field equations write
\begin{equation}\label{H2}
H^2=\frac{k}{3}(\rho_m+\rho_r+\rho_d)
\end{equation}
\begin{equation}\label{rhomd}
\dot \rho_m+3H\rho_m=0
\end{equation}
\begin{equation}\label{rhord}
\dot\rho_r+4H\rho_r=Q
\end{equation}
\begin{equation}\label{rhodd}
\dot\rho_d+3(1+w)H\rho_d=-Q
\end{equation}
$\rho_m$, $\rho_r$ and $\rho_d$ are respectively the densities of matter, radiation and vacuum. A dot means a derivative with respect to proper time $t$. We consider a vacuum dark energy defined by $w=-1$ and assume that $Q>0$. This implies that the vacuum energy is cast into radiation. We define the following dimensionless variables
\begin{equation}\label{y1}
y_1=\frac{k}{3}\frac{\rho_m}{H^2}
\end{equation}
\begin{equation}\label{y2}
y_2=\frac{k}{3}\frac{\rho_r}{H^2}
\end{equation}
\begin{equation}
y_3=\frac{k}{3}\frac{\rho_d}{H^2}
\end{equation}
\begin{equation}
q=\frac{k}{3}\frac{Q}{H^3}
\end{equation}
The values of $y_1$, $y_2$ and $y_3$ today are respectively the present values of the density parameters for matter $\Omega_{m0}$, radiation $\Omega_{r0}$ and dark energy $\Omega_{d0}$. The field equations rewrite
\begin{equation}\label{eq1}
y_1'=y_1(-3+3y_1+4y_2)
\end{equation}
\begin{equation}\label{eq2}
y_2'=y_2(-4+3y_1+4y_2)+q
\end{equation}
with the constraint $1=y_1+y_2+y_3$. A prime means a derivative with respect to $N=\ln a$ with $a$ the scale factor of the FLRW metric. We assume that Universe is expanding, i.e. $H>0$. Then $q>0$ since $Q>0$ and $a$ increases with time. $N$ is thus a time variable. Moreover, $q$, $y_1$ and $y_2$ are, at least formally, some functions of the redshift $z$. It follows that $q=q(z)=q(y_1,y_2)$. Hence the dynamical system (\ref{eq1}-\ref{eq2}) is autonomous. However, the variables $y_i$ are not normalised. $y_1$ is always positive since $\rho_m= \rho_{m0}(1+z)^3$ and the energy density of matter today, $\rho_{m0}$, is positive. But $y_2$ and/or $y_3$ can be negative if the radiation and/or vacuum energy densities are negative (see section \ref{s3} for a discussion about negative energy density).\\
In the following, we use the tools of phase space analysis to study the properties of the above defined theory. Our goal is not to make a full dynamical analysis of the above system. It is to find the conditions that allow it to reproduce the full dynamics of our Universe from early time inflation to late time $\Lambda CDM$ acceleration.
\subsection{Equilibrium points for finite values of $y_1$ and $y_2$} \label{s12}
\begin{figure}[h]
\centering
\includegraphics[width=6cm]{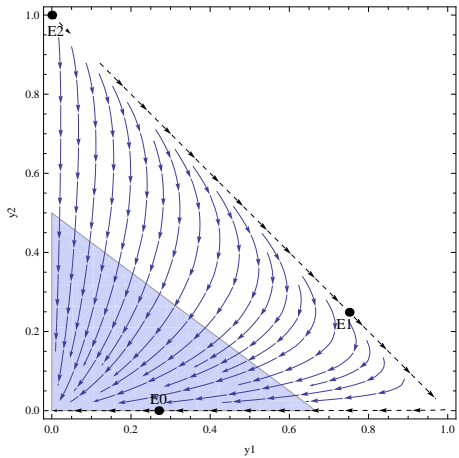}
\caption{\scriptsize{\label{fig1} Phase space of the $\Lambda CDM$ model. The dashed trajectory corresponds to our Universe and the points to Universe today ($E_0$), at the $CMB$ ($E_1$) and nucleosynthesis ($E_2$) times. The gray area is the part of the phase space where Universe expansion is accelerated.}}
\end{figure}

We first examine what are the equilibrium points of this system when the $y_i$ are finite. There could be some equilibrium points for diverging values of $y_1$ and $y_2$. However, we will not need to look for them to find how the system can describe an inflation period, following by a $\Lambda CDM$ behaviour. The equilibrium points for finite values of $y_1$ and $y_2$ are defined by
\begin{itemize}
\item the set of points $P$ such that $(y_1,y_2)=(\frac{1}{3}(3-4q(y_1,y_2)),q(y_1,y_2))$
\item the set of points $P_\pm$ such that $(y_1,y_2)=(0,\frac{1}{2}(1\pm\sqrt{1-q(0,y_2)}))$
\end{itemize}
These sets of equilibrium points do not always exist. For instance, $P_\pm$ points are complex when $q(0,y_2)>1$.\\
When $q=0$, we recover the $\Lambda CDM$ model with its equilibrium points: a source in $(y_1,y_2)=(0,1)$ (radiation domination), a saddle in $(1,0)$ (matter domination) and a sink in $(0,0)$ (vacuum domination). The phase space $(y_1,y_2)$ for the $\Lambda CDM$ model is plotted on figure \ref{fig1}. The dashed trajectory represents Universe as we observe it, i.e defined by the density parameters today for matter and radiation, $\Omega_{m0}=0.27$ and $\Omega_{r0}=8.27\times 10^{-5}$. On this trajectory, there are three important events $E_0$, $E_1$ and $E_2$. They correspond respectively to Universe today ($z=0$), at the CMB ($z=1080$) and nucleosynthesis ($z=4.25\times 10^{8}$) epochs. Figure \ref{fig1} also offers a nice geometrical illustration of the coincidence problem from dynamical system viewpoint. The coincidence problem\cite{Gar00} consists in the fact that at our present time, whereas $y_2$ is very small, $y_1$ is of the same order as $y_3$. This selects, as our Universe history, the dashed trajectory of figure \ref{fig1} that follows so closely (coincidentally) the limit of the triangular domain where both $\rho_r$ and $\rho_d$ are positive.\\
In the rest of the paper we are going to consider that $q\not =0$.
\subsection{Conditions for an early accelerated expansion} \label{s13}
The $\Lambda CDM$ model does not contain any inflation epoch. All its trajectories start with a decelerated expansion in $(y_1,y_2)=(0,1)$ during which Universe is radiation dominated. Things can be different when $q\not =0$. Let us show it by looking for the points $(y_1,y_2)$ describing an early accelerated expansion. They are such that $d^2 a/dt^2>0$ or equivalently
\begin{equation}\label{eq3}
1-\frac{3}{2}y_1-2y_2>0
\end{equation}
This equation implies that the only set of equilibrium points with finite values of $(y_1,y_2)$ in agreement with an accelerated expansion is the set $P_-$. But can some points of this set be sources? Indeed, $P_-$ points are real if $0\leq q\leq 1$. It follows that for a point of $P_-$, $0\leq y_2\leq 1/2$, and since $y_1\geq 0$, $y_1'\rightarrow 0^-$. Hence, any trajectory starting from a point of $P_-$ where $y_1=0$ is such that $y_1'<0$ and thus such that $y_1<0$ in the neighbourhood of this point, contradicting the inequality $y_1\geq 0$. There is thus no source point in the $P_-$ set describing an accelerated expansion with finite values of $y_1$ and $y_2$. Consequently, if Universe starts by an inflation epoch, it occurs for some diverging values of $y_1$ and/or $y_2$. Equation (\ref{eq3}) indicates that, at least, $y_2$ should diverge negatively since $y_1>0$. A negative radiation energy density is thus necessary to trigger inflation.
\subsection{Finite scale factor singularity} \label{s14}
In appendix \ref{a1}, we show that a singularity in agreement with an early time inflation always starts at a finite value $N_s$ of $N$ and never in $N\rightarrow -\infty$. Let us summarise this proof. We begin to assume that Universe starts in $N\rightarrow -\infty$. Then the possible early time behaviours of $y_1$ and $\rho_m/\rho_r$ in agreement with inflation are such that $H^2<<\rho_m<<\mid\rho_r\mid$. However, if we examine the possible behaviours of $\rho_r$ in $N\rightarrow -\infty$, we find that they disagree with the previous inequalities. We thus conclude that an early time inflation implies a finite scale factor singularity in $N=N_s$.\\\
In the next subsection, we present a class of theories describing a Universe that begins by an accelerated expansion and tends at late time to the $\Lambda CDM$ model.
\subsection{A coupling function of the form $q=q_0 y_1^m+q_1 y_2^n$} \label{s15}
\begin{figure}[h]
\centering
\includegraphics[width=6cm]{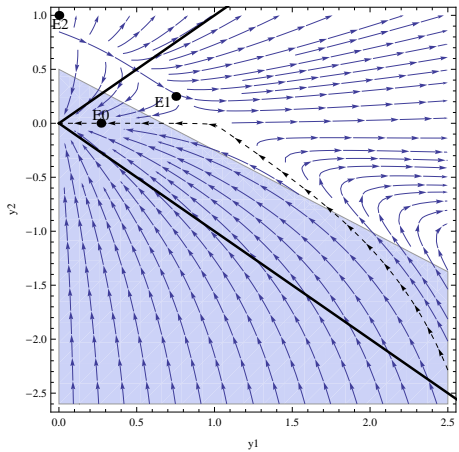}
\caption{\scriptsize{\label{fig2} Phase space of the vacuum/radiation coupled model with $q=0.01y_1^4+y_2^4$. The dashed trajectory contains the point $E_0$ for our Universe today. It misses the point $E_1$ (CMB) and $E_2$ (nucleosynthesis) of the $\Lambda CDM$ model. The black lines define the limit $\rho_m=\mid \rho_r\mid$.}}
\end{figure}

\begin{figure}[h]
\centering
\includegraphics[width=6cm]{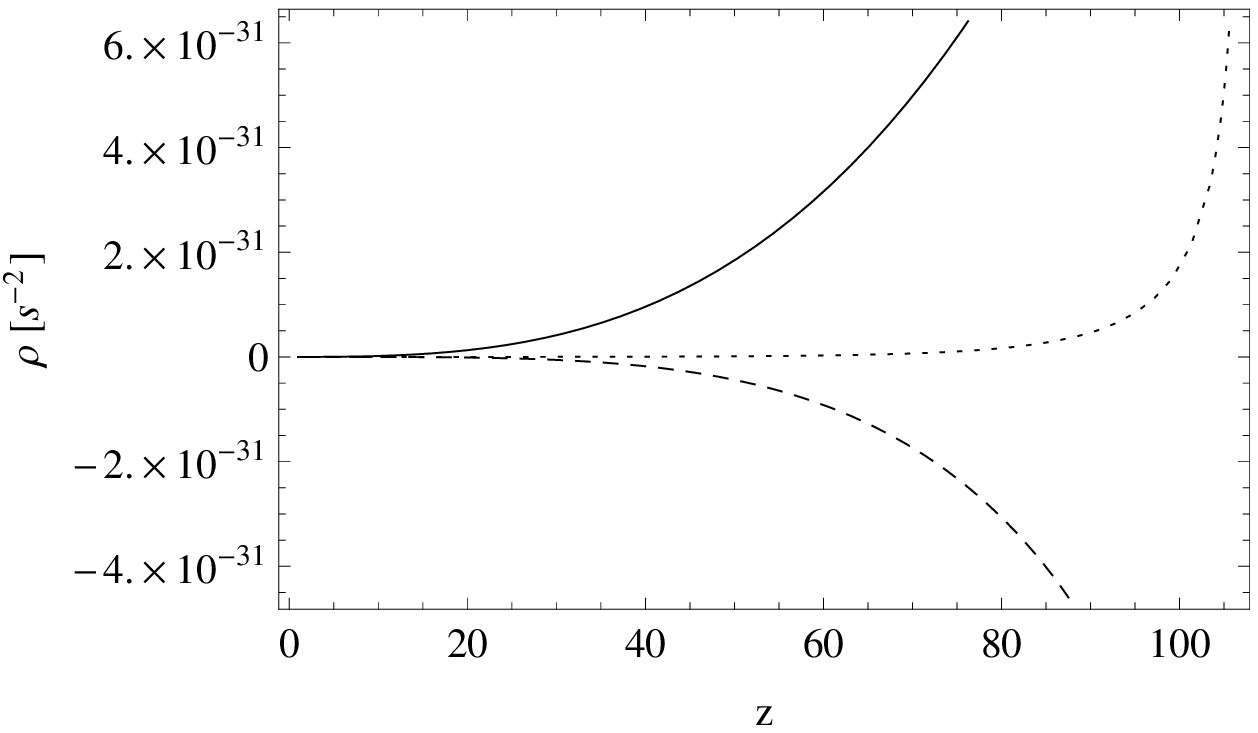}
\includegraphics[width=6cm]{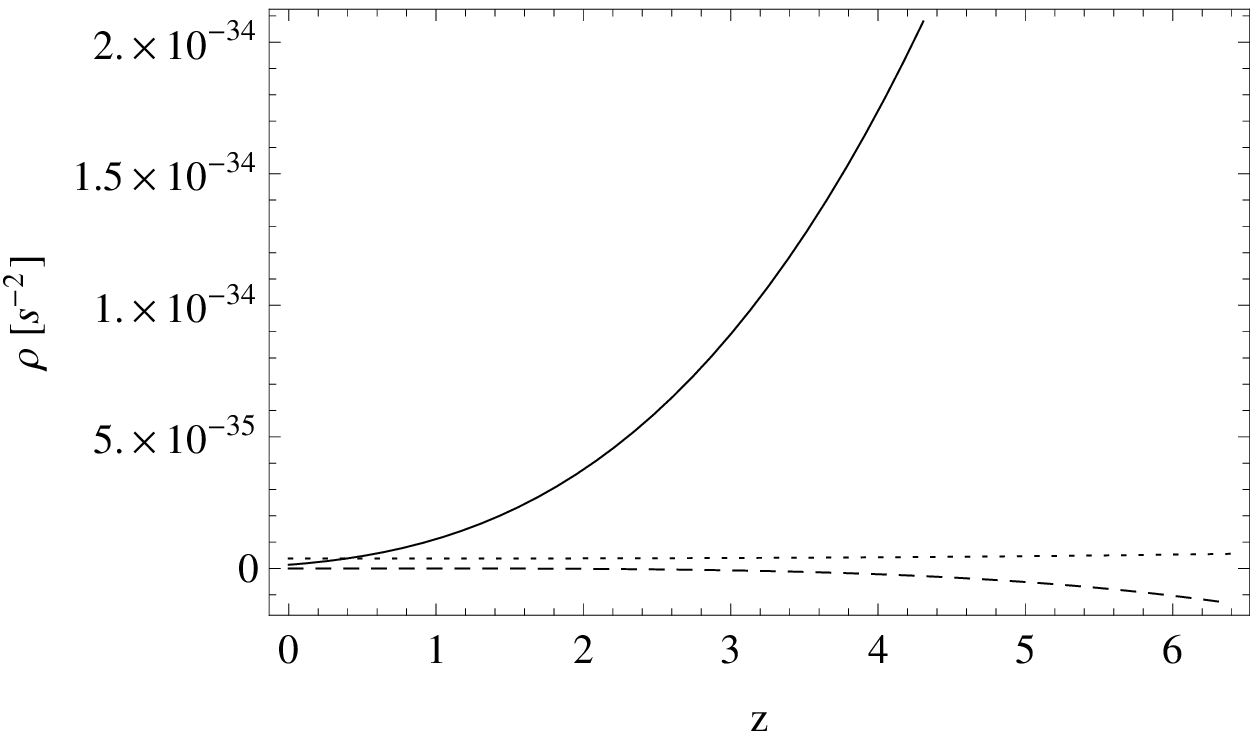}
\caption{\scriptsize{\label{fig3}Evolution of the matter, radiation (dashed) and vacuum (dotted) densities when $q=0.01y_1^4+y_2^4$, $H_0=70 km/s/Mpc$, $\Omega_{m0}=0.27$ and $\Omega_{r0}=8.27\times 10^{-5}$. The first graph shows that the singularity occurs in redshift $z=106$. The second graph shows the late time convergence to the $\Lambda CDM$ model for which $\rho_d$ is a constant.}}
\end{figure}

We consider the class of theories defined by
$$
q=q_0 y_1^m+q_1 y_2^n
$$
We want that this form of coupling leads to a Universe that starts by an inflation phase and tends to a $\Lambda CDM$ model. This last requirement means that at late time $q(0,0)\rightarrow 0$ and thus $m>0$ and $n>0$. The larger $m$ and $n$, the faster the model converges to the late time equilibrium point $(y_1,y_2)=(0,0)$ and approaches the $\Lambda CDM$ model. Moreover, since at early time we must have $y_2<0$ to trigger inflation, $n$ have to be an integer such that $q$ be real. Considering these properties for $m$ and $n$, we perform some numerical simulations for various values of these parameters. They lead us to consider the early time inequalities $y_1<<y_2$ and $y_1^m<<y_2^n$ for which we get the asymptotical behaviours
$$
y_1\rightarrow \beta e^{-\frac{4 y_2^{2-n}}{(-2+n) q_1}}
$$
$$
y_2\rightarrow -\left[(n-1) (N q_1+\alpha)\right]^{\frac{1}{1-n}}
$$
$\alpha>0$ and $\beta>0$ are some integration constants. Other inequalities between $y_1$ and $y_2$ always lead to some decelerated expansion or inconsistencies. For instance, if we consider $y_1<<y_2$ and $y_1^m>>y_2^n$, the field equations imply that $y_2$ is finite at early times and there is thus no inflation.\\
The above asymptotical behaviours for $y_1$ and $y_2$ show that when $n$ is an integer larger than $1$ (which is necessary if one wishes to converge quickly to the $\Lambda CDM$ model), $y_2$ diverges in $N=-\alpha q_1^{-1}<0$ while $y_1(N_s)$ is a constant. A singularity thus occurs in $N_s$ for which the scale factor $a= e^{N_s}$, the matter density $\rho_m(N_s)= e^{-3N_s}$ and the Hubble function $H(N_s)=\sqrt{\rho_m(N_s)/y_1(N_s)}$ are some non vanishing constants while the densities $\rho_r(N_s)\rightarrow -\infty$ and $\rho_d(N_s)\rightarrow \infty$ (since $1=y_1+y_2+y_3$). Since we choose $q(y_1(N_s),-\infty)> 0$, it implies that when $q_1>0$ ($q_1<0$), $n$ has to be an even (respectively an odd) integer.\\
\\
A numerical example of the above results is plotted on figure \ref{fig2} when $q=0.01y_1^4+y_2^4$. The dashed trajectory corresponds to a Universe with the observed values $\Omega_{m0}=0.27$ and $\Omega_{r0}=8.27\times 10^{-5}$ today. Evolutions of matter, radiation and vacuum densities are plotted on figure \ref{fig3}.
Let us explain what happens physically on this trajectory. Universe starts by a finite scale factor singularity (see first graph on figure \ref{fig3}) in $N=-4.68$. The radiation energy density is then diverging and negative ($y_2<0$), $\rho_m<<\rho_r$, whereas vacuum energy density is diverging and positive ($y_3>0$). Universe expansion accelerates: we are in the inflationary era. Because of the positive coupling $q>0$, vacuum decays into radiation. Consequently, the absolute values of their densities become smaller: $\rho_d>0$ decreases and $\rho_r<0$ increases. When matter dominates sufficiently (i.e. with respect to equation (\ref{eq3})), the expansion naturally decelerates. When both the radiation and vacuum energy densities become negligible, $q$ is small and Universe begins to approach the $\Lambda CDM$ model (well after the $\Lambda CDM$ CMB epoch in $E_1$ when $y_2$ is positive and not negligible). The vacuum energy has not completely decayed and now stays nearly constant (see second graph on figure \ref{fig3}) thanks to the smallness of $q$. As usual for the $\Lambda CDM$ model, both matter and radiation densities decrease until vacuum dominates anew and triggers a second acceleration of the expansion. To realise this scenario, numerical simulations show that $q_0$ have not to be too small ($q_0>10^{-4}$) otherwise energy exchange between vacuum and radiation near the matter dominated era is not sufficient to allow the radiation density to become positive at present time (or from a dynamical system viewpoint, the $\Lambda CDM$ trajectory in the phase space is approached from below with $y_2<0$ well after $y_1<1$). Moreover, the largest $q_0$, the largest should be $m$ to allow a quick convergence to the $\Lambda CDM$ model at late time, in particular when $y_1< 1$. 
\\
\\
Although this scenario reproduces two accelerated epochs for the expansion, it is obviously not in agreement with observations. It joins the $\Lambda CDM$ model only at the matter epoch and we have to consider negative energy density as soon as a redshift $z\propto 1$. Moreover, the singularity occurs around $z=106$ ! This does not prove that we cannot find a function $q$ able to reproduce observations. However, it is likely that such a function should be more elaborate than the simple model of this section. For instance, it should allow to cross the line $y_2=0$ far from the matter dominated point, i.e. with $y_1>1$ to avoid that the trajectory be immediately attracted by the De Sitter attractor in $(y_1,y_2)=(0,0)$. Vacuum energy density would thus be negative before the trajectory potentially turn back to join the $\Lambda CDM$ trajectory, at least in $E_1$.\\\\
This difficulty to reach the $\Lambda CDM$ model early in the Universe evolution mainly comes from the fact that inflation implies $y_2<0$. Indeed, if we look at the phase space on figure \ref{fig2}, we remark that reaching the nucleosynthesis point $E_2$ of the $\Lambda CDM$ model would be more natural if Universe could start its evolution on the left of the line $y_2=0$, i.e. with $y_1<0$. Such a requirement is satisfied in the next section where we consider a coupling between vacuum and matter.
\section{Vacuum energy coupled to matter} \label{s2}
We now consider a coupling between vacuum and matter such as vacuum is cast into matter.
\subsection{Field equation} \label{s21}
When dark energy is coupled to matter, the field equations write
\begin{equation}
H^2=\frac{k}{3}(\rho_m+\rho_r+\rho_d)
\end{equation}
\begin{equation}\label{rhomd1}
\dot\rho_m+3H\rho_m=Q
\end{equation}
\begin{equation}\label{rhord1}
\dot\rho_r+4H\rho_r=0
\end{equation}
\begin{equation}\label{rhodd1}
\dot\rho_d+3(1+w)H\rho_d=-Q
\end{equation}
We still consider a vacuum energy with $w=-1$ and $Q>0$. Note that in \cite{Pav09}, this last inequality is required for respect of the second law of thermodynamics. Using the same variables as in section \ref{s1}, we get
\begin{equation}\label{eq4}
y_1'=y_1(-3+3y_1+4y_2)+q
\end{equation}
\begin{equation}\label{eq5}
y_2'=y_2(-4+3y_1+4y_2)
\end{equation}
with the constraint $y_1+y_2+y_3=1$. $y_1$ and $y_3$ can be negative but not $y_2$ since now $\rho_r=\rho_{r0}(1+z)^4>0$ with $\rho_{r0}$ the positive radiation energy density today. We still assume an expanding Universe with $H>0$.
\subsection{Equilibrium points for finite values of $y_1$ and $y_2$ and accelerated expansion} \label{s22}
The sets of equilibrium points for finite values of $y_1$ and $y_2$ are defined by
\begin{itemize}
\item the set $P$ such that $(y_1,y_2)=(-q(y_1,y_2),\frac{1}{4}(4+3q(y_1,y_2)))$
\item the set $P_\pm$ such that $(y_1,y_2)=(\frac{1}{6}(3\pm\sqrt{3}\sqrt{3-4q(y_1,0)}),0)$
\end{itemize}
The condition for an accelerated expansion is still given by equation (\ref{eq3}). It follows that only the $P_\pm$ points could correspond to an accelerated expansion if respectively $\frac{1}{4} \left(1\mp\sqrt{3} \sqrt{3-4 q}\right)>0$. Once again, we are going to show that they cannot be sources. Both sets of points are such that $y_2=0$ and are real if $0\leq q\leq 3/4$. For the set of points $P_-$, it means that $0\leq y_1\leq 1/2$ and thus $y_2'<0$. It follows that any trajectory starting from a point of $P_-$ is such that $y_2'<0$ and thus such that $y_2<0$ in the neighbourhood of this point, contradicting the inequality $y_2\geq 0$. Hence, there is no source point in $P_-$ describing an accelerated expansion. In the same way, for the set of points $P_+$, we have $1/2\leq y_1\leq 1$ and thus $y_2'<0$. So, here also, the set $P_+$ cannot contain any source point without contradicting the fact that $y_2\geq 0$. Hence, no equilibrium point with finite values of $y_1$ and $y_2$ can be the source of an accelerated expansion. Such a behaviour thus takes place for a divergent and negative value of $y_1$ (a positive and divergent value of $y_1$ with $y_2>0$ is not in agreement with an accelerated expansion as shown by equation (\ref{eq3})). This implies that a negative energy density for matter is necessary to trigger inflation. If one looks at the phase space of the $\Lambda CDM$ model on figure \ref{fig1}, it means that we can have some trajectories describing an early time inflation and coming from the left side of the $\Lambda CDM$ $E_2$ point. They could thus converge easily to the $\Lambda CDM$ trajectory as soon as the nucleosynthesis epoch. We illustrate these results in subsection \ref{s24}.
\subsection{Finite scale factor singularity} \label{s23}
A singularity in agreement with an early time inflation takes place at a finite value of $N=N_s$ but in $N\rightarrow -\infty$ when $\rho_m\rightarrow -4/3\rho_r<0$. The proof of this statement is given in appendix \ref{a2}. We summarise it in this subsection. In this proof, we start by studying the possible asymptotical behaviours of $y_2$ in $N\rightarrow -\infty$ during an inflation period. We conclude that $y_2$ must tend to $+\infty$ and thus $H^2<<\rho_r$. Then, we examined the quantity $\rho_r/\rho_m$. We show that when it tends to $-3/4$, an inflation period can take place in $N\rightarrow -\infty$. Otherwise, it has to tend to $0^+$, implying that inflation could also occur in $N\rightarrow -\infty$ if $H^2<<\rho_r<<\mid\rho_m\mid$. But examining the possible behaviours of $\rho_m$ in agreement with an inflation epoch in $N\rightarrow -\infty$, we show that none of them is compatible with the previous inequalities.\\
We thus conclude that an early time inflation implies a finite scale factor singularity in $N=N_s$ but when $\rho_m\rightarrow -4/3\rho_r$. When the singularity arises in $N=N_s$, $\rho_r$ is finite whereas $\rho_m$ and $\rho_d$ can diverge.\\\\
In the next subsection, we present a class of theories describing a Universe that begins by an accelerated expansion and then joins the $\Lambda CDM$ model at early time.
\subsection{A coupling function of the form $q=q_1 y_2^n$} \label{s24}
\begin{figure}[h]
\centering
\includegraphics[width=6cm]{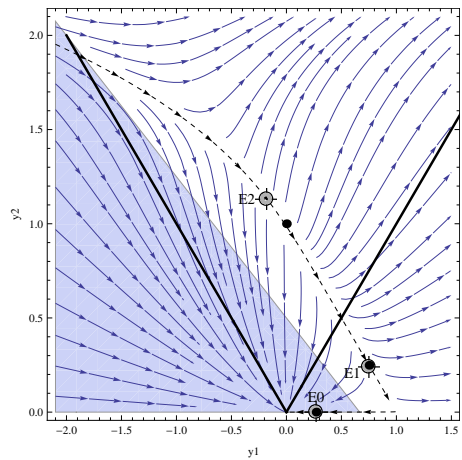}
\caption{\scriptsize{\label{fig4} Phase space of the vacuum/matter coupled model with $q=0.07y_2^8$. The dashed trajectory contains our Universe today and thus the point $E_0$ with in $z_0=0$, $\Omega_{m0}=0.27$ and $\Omega_{r0}=8.27\times 10^{-5}$. The two other black points are the CMB and nucleosynthesis points of the $\Lambda CDM$ model. The gray points $E_1$ and $E_2$ are the location on the dashed trajectories of the events at redshift $z=1080$ and $z=4.25\times 10^8$. The black lines define the limit $\rho_m=\mid \rho_r\mid$.}}
\end{figure}

\begin{figure}[h]
\centering
\includegraphics[width=6cm]{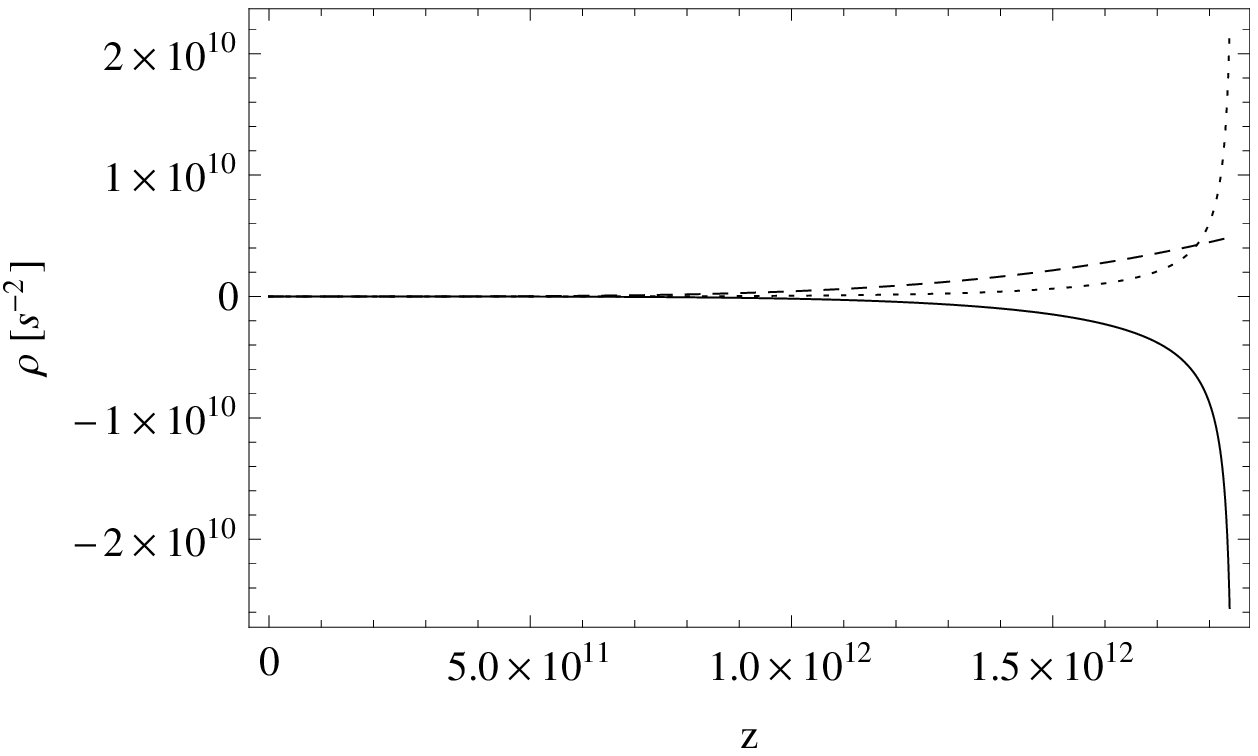}
\includegraphics[width=6cm]{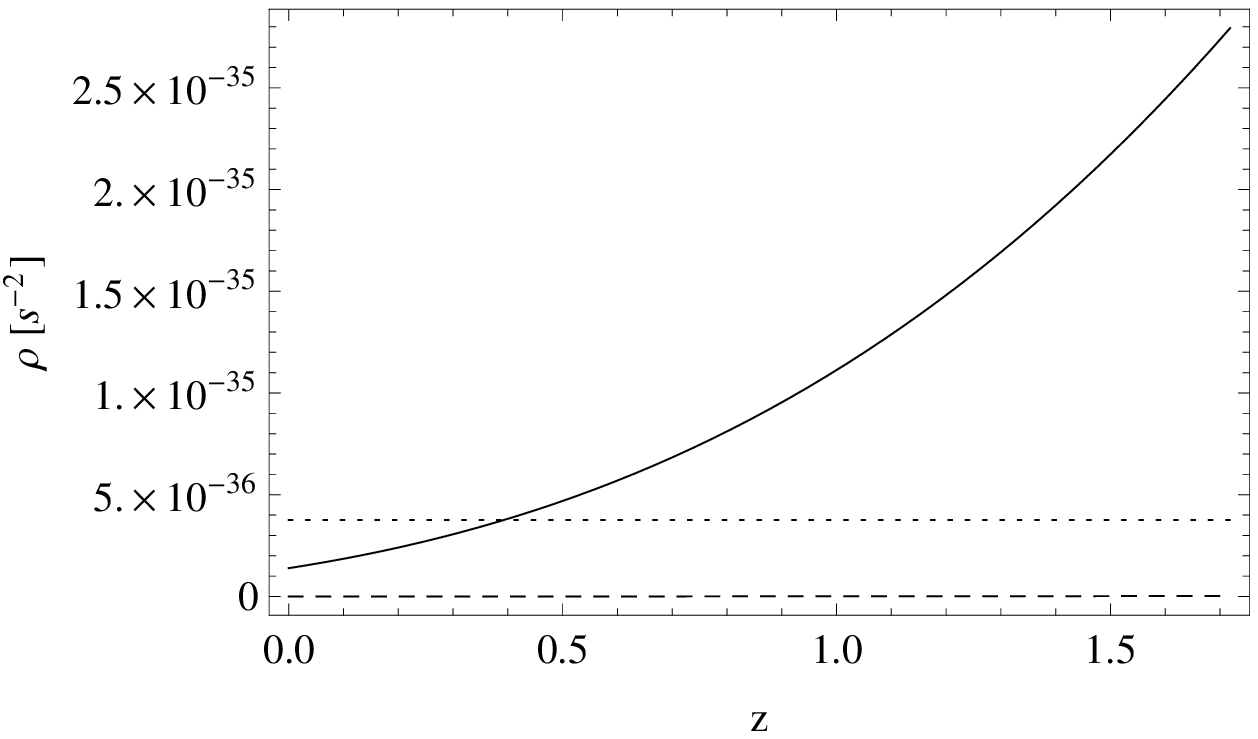}
\caption{\scriptsize{\label{fig5} Evolution of the matter, radiation (dashed) and vacuum (dotted) densities when $q=0.07y_2^8$, $H_0=70 km/s/Mpc$, $\Omega_{m0}=0.27$ and $\Omega_{r0}=8.27\times 10^{-5}$. The first graph shows that the singularity occurs in $z=1.84\times 10^{12}$. The second graph shows the convergence to the $\Lambda CDM$ model for which $\rho_d$ is a constant.}}
\end{figure}

We consider the coupling function
$$
q=q_1 y_2^n
$$
We want that such a coupling allows an inflation period at early time and a fast convergence to the $\Lambda CDM$ model. These properties imply some constraints on $q_1$ and $n$. Since $y_2>0$ and $q>0$, $q_1>0$. We also have $n>0$ such that $q\rightarrow 0$ when $y_2\rightarrow 0$ near the de Sitter attractor of the $\Lambda CDM$ model. For an early time approach of the $\Lambda CDM$ model, $q$ should be small in the neighbourhood of the radiation source point $(y_1,y_2)=(0,1)$ of this last model. This will always be the case for large values of $n$ since as soon as $y_2$ becomes smaller than $1$, $q$ is small if $n$ is large. It follows that we can have large values of $q_1$ if we have large values of $n$. However, $q_1$ and $n$ should not be too small together, otherwise vacuum would not exchange enough energy with matter before the coupling $q$ becomes small and consequently matter density would always stay negative.\\
A numerical example with the above properties is given by $q=0.07 y_2^8$ and plotted on figure \ref{fig4}. The dashed trajectory corresponds to a Universe with $\Omega_{m0}=0.27$ and $\Omega_{r0}=8.27\times 10^{-5}$ today. The matter, radiation and vacuum densities are plotted on figure \ref{fig5}. As shown by figure \ref{fig4}, such a model is able to describe an early time inflation phase for a Universe converging to the $\Lambda CDM$ model as early as the nucleosynthesis epoch. Numerical simulations indicate that at early time, during inflation, $y_2<<y_1$ but $y_2^n>>y_1$. Without specifying the values of $q_1$ and $n$, these inequalities lead to the following asymptotical behaviours for $y_1$ and $y_2$:
$$
y_1\rightarrow  -\frac{\sqrt{2 q_1 y_2^n-\alpha y_2^2(3n-6)}}{\sqrt{3} \sqrt{-2+n}}
$$
with $\alpha$ an integration constant. Hence for $n>2$, $y_1\propto -\sqrt{y_2^n}$ at early times and considering equation (\ref{eq5}), it comes
$$
y_2\rightarrow  \left[n/2 \left(-\frac{\sqrt{6} N \sqrt{q_1}}{\sqrt{-2+n}}-\beta \right)\right]^{-2/n}
$$
with $\beta$ an integration constant. A finite scale factor singularity occurs in $N_s=-\frac{\beta\sqrt{-2+n}}{\sqrt{6q_1}}<0$. Then, inflation starts with $y_2\rightarrow \infty$ and $y_1\rightarrow -\infty$, in agreement with the results of subsection \ref{s22} for this last quantity. We also have $\rho_r(N_s)\rightarrow e^{-4N_s}$, $H(N_s)=0$ (since $y_2(N_s)\propto \rho_r/H^2$ diverges\footnote{Note that $H(N_s)=0$ does not mean that the scale factor tends to a constant since $\dot H(N_s)\not =0$. Hence it does not contradict an inflation behaviour.}), $\rho_m(N_s)\rightarrow -\infty$ (since $y_1/y_2=\rho_m/\rho_r$ diverges in $N=N_s$ when $n>2$) and $\rho_d(N_s)\rightarrow \infty$ (since $H^2\propto \rho_r+\rho_m+\rho_d=0$ in $N=N_s$). The early time behaviours for the densities are shown on the first graph of figure \ref{fig5}.\\
Let us explain what happens physically. Universe starts by a finite scale factor singularity. Vacuum and matter densities diverge but not the radiation density that is a constant (see first graph on figure \ref{fig5}). We have then $y_1\rightarrow -\infty$ with $y_1>>y_2$ and $y_3\rightarrow +\infty$. Universe expansion is thus accelerated as shown by equation (\ref{eq3}). Vacuum and matter dominate Universe content. But vacuum exchanges its energy with matter and both of them become (in absolute value) smaller when time increases: $\rho_d>0$ decreases and $\rho_m<0$ increases. When these densities are small enough, radiation dominates and Universe expansion begins to decelerate. $q$ becomes smaller and smaller as $y_2$ decreases and we approach a $\Lambda CDM$ model. Vacuum has not completely decayed and its energy density is now constant (see second graph on figure \ref{fig5}). At late time, it dominates Universe dynamics anew and a second pahse of accelerated expansion begins.\\
A fundamental difference with respect to the case for which vacuum is coupled to radiation is that now, a trajectory of the phase space $(y_1,y_2)$ can join from the left ($y_1<0$) the trajectory of the $\Lambda CDM$ model as soon as the early radiation epoch (i.e. the neighbourhood of the $\Lambda CDM$ source point $(y_1,y_2)=(0,1)$). However, as shown on figure \ref{fig4}, there are some differences with the $\Lambda CDM$ model. Although the trajectories of the coupled and $\Lambda CDM$ models are very similar between the points $(y_1,y_2)=(0,1)$ and $(0,0)$ (the sink point corresponding to a De Sitter Universe), their time parameters are different. More specifically, for a point $(y_1,y_2)$ of the $\Lambda CDM$ trajectory corresponding to a redshift $z_1$, the same point of the coupled model corresponds to a redshift $z_2<z1$. Hence, if one considers that the CMB redshift is $z=1080$, then the CMB epoch occurs for a ratio $y_2/y_1=\rho_r/\rho_m$ slightly larger in the coupled model than in the $\Lambda CDM$ model. Or if we consider that the CMB epochs should occur for the same ratio $y_2/y_1=\rho_r/\rho_m$ in both models, then the CMB epoch occurs at a redshift slightly smaller in the coupled model than in the $\Lambda CDM$ model. These differences should be observable on the CMB\cite{Mor13} or during nucleosynthesis time. Note also that when taking $q=0.07y_2^8$, $H_0=70 km/s/Mpc$, $\Omega_{m0}=0.27$ and $\Omega_{r0}=8.27\times 10^{-5}$ for present time, one finds for this particular model that $\Omega_d(z=1100)=1.23\times 10^{-7}$ and $\Omega_d/\Omega_r(z=4.25\times 10^8)=0.042$. These last values are respectively in agreement with Planck data\cite{Ade13} and big-bang nucleosynthesis\cite{Wri07}.\\
\section{Discussion}\label{s3}
In this paper we consider General Relativity with matter, radiation and dark energy. We assume that dark energy is a decaying vacuum coupled to matter or radiation, that the coupling function $Q$ is positive and Universe is expanding. Then we look for the properties of this theory such as Universe starts with an inflation phase and converges to a $\Lambda CDM$ model. We showed that this scenario is possible only if the fluid coupled to dark energy have a negative energy density at early time. Moreover, Universe generally starts by a finite scale factor singularity but when matter is coupled to dark energy and behaves at early time as a negative radiation density such that $\rho_m\rightarrow -4/3\rho_r<0$. Let us discuss some of these properties.\\\\
This is not the first time that a finite scale factor singularity appears when trying to unite cosmological epochs. For instance, it arises when one uses scalar-tensor theories\cite{Hry10}. However, as shown in the appendix, finite scale factor singularities can be avoided if during inflation $\rho_m/\rho_r$ tends to $-4/3$ when matter is coupled to dark energy or by allowing negative $Q$ when radiation is coupled to dark energy, a case that we exclude in this paper where we consider $Q>0$.\\\\
Let us also discuss the presence of negative energy density to trigger inflation. A first point of view about negative energy density is to consider that, because it occurs at early times, it could be explained or even cured by a quantum theory of gravity\cite{Kuo97}. There is no real consensus about the existence of negative energy density. It is sometimes used to build traversable wormhole whereas some uncertainty-principle-type limitation about the magnitude and duration of a negative energy density has been calculated under some conditions in \cite{For97}\cite{Pfe98}. 
A second point of view could be to consider that the two coupled fluids make one fluid of density $\rho$, hoping that it stays positive during the inflation period. Hence, let us call $y_i$ the fluid coupled to dark energy $y_3$, and $y_j$ the non coupled fluid. Then the equation of state $w$ for $\rho$ would be $w+1\propto y_i(y_i+y_3)^{-1}$. Since $\rho$ has the sign of $y_i+y_3$ and we have the constraint equation $1=y_i+y_j+y_3$, $\rho$ is positive at early time during inflation when $y_j<1$ (like in section \ref{s15} when $j=1$ and the integration constant $\beta<1$). Then $w<-1$ since $y_i<0$ and $\rho$ is positive but is a ghost dark energy. However, if $y_j>1$ (like in section \ref{s24} when $j=2$), $\rho$ is negative but we have a quintessent dark energy with $w>-1$. Note that when $y_i=y_1$, this point of view is equivalent to realise unification of inflation, dark matter and dark energy into a single field\cite{Arb09}, an idea coming from the string landscape\cite{Lid06}.\\
A last idea to avoid the necessity of a negative energy density to trigger inflation would be to consider that both radiation and matter are coupled with vacuum. The advantage is that the divergence of $y_1$ and $y_2$ can be avoided at early time during inflation by choosing an appropriate form for $q$, keeping matter and radiation densities positive and even finite. This would then also allow the domination of a vacuum energy with a constant energy density at early as well as at late time. Such a scenario is described in \cite{Lim12} where the vacuum energy density is modelised as a series of even power of the Hubble function. It suggests that some scalar-tensor theories can be in agreement with an accelerated expansion at early and late time\cite{Wan13}. Indeed, when expressed in the Einstein frame, a scalar-tensor theories is cast into General Relativity with matter and radiation coupled in the same way to the scalar field. If some accelerated expansion phases arise at both early and late time in the Einstein frame, they can be (not always) conserved by the conformal transformation in the Brans-Dicke frame. This will be the subject of future work.
\appendix
\section{Finite scale factor singularity when dark energy is coupled to radiation} \label{a1}
We want to show that an inflation epoch cannot take place in $N\rightarrow -\infty$. For that, we assume the opposite and proceed in two main steps. In the first one, we assume that $y_1$ and $\rho_m/\rho_r$ tend to some constants or diverge in $N\rightarrow -\infty$. We show that inflation occurs in $N\rightarrow -\infty$ only if then $H^2<<\rho_m<<\mid\rho_r\mid$. But considering the possible behaviours of $\rho_r$ with  respect to these inequalities, we show that none of them agrees with an inflation epoch at that time. In a second step, we assume that $y_1$ and $\rho_m/\rho_r$ do not tend to some constants or diverge in $N\rightarrow -\infty$, i.e. they are thus oscillating. Then, we show that these oscillations are not in agreement with an inflation epoch in $N\rightarrow -\infty$.\\
Let us begin by considering that $y_1$ and $\rho_m/\rho_r$ asymptotically diverge or tend to some constants in $N\rightarrow -\infty$.
\begin{itemize}
\item $y_1$ cannot tend to a non vanishing constant when $N\rightarrow -\infty$ because, from (\ref{y1}), it would follow that $H^2\rightarrow \rho_m$ and there would thus be no inflation.
\item $y_1$ cannot tend to a vanishing constant when $N\rightarrow -\infty$. Indeed, since we assume an early time inflation epoch, we must also have $y_2<0$ (see section \ref{s13}). Then it would follow from (\ref{eq1}) that $y_1'<0$. But $y_1$ cannot approach $y_1=0$ in $N\rightarrow -\infty$ with $y_1'<0$ and stay positive. This case is thus excluded.
\item The two previous points imply that the only behaviour of $y_1$ in agreement with an inflation period in $N\rightarrow -\infty$ is $y_1\rightarrow +\infty$ and thus $\rho_m>>H^2$. We then examine the possible behaviours of $\rho_m$ with respect to $\rho_r$.
\begin{itemize}
\item $\rho_m/\rho_r$ cannot be such that $\rho_m/\rho_r\rightarrow \alpha+\epsilon_0(N)$ in\footnote{We need to consider a small perturbation $\epsilon_0(N)$ because its integral intervenes in the special case $\alpha =-4/3$.} $N\rightarrow -\infty$ with $\alpha$ a non vanishing and negative (since $\rho_m$ and $\rho_r$ have opposite signs) constant and $\epsilon_0(N)<<\alpha$. Indeed, in this case, we would have $y_1\rightarrow \alpha y_2+\epsilon(N)$ with $\epsilon(N)<<y_2$. When $\alpha\not =-4/3$ and since $y_2$ diverges during inflation, it comes from (\ref{eq1}) that $1/y_1\simeq - \frac{3\alpha+4}{\alpha}N$. But then, when $N\rightarrow -\infty$, $y_1$ and thus $y_2$ tend to vanish that disagrees with inflation. If now $\alpha =-4/3$, from (\ref{eq1}) we deduce that $y_1\propto e^{-3N+3\int\epsilon dN}>0$ and thus $y_2\propto-3/4e^{-3N+3\int\epsilon dN}<0$. But introducing these quantities in (\ref{eq2}), one shows that $q<0$ that disagrees with our assumption $q>0$. 
\item $\rho_m/\rho_r$ cannot be such that $\rho_m/\mid \rho_r\mid\rightarrow +\infty$. Then $y_1>>\mid y_2\mid$ and since we show above that $y_1\rightarrow +\infty$, (\ref{eq3}) implies that there is no inflation.
\item From the two previous points, the only possibility for $\rho_m$ in $N\rightarrow -\infty$ is thus $\rho_m/\mid \rho_r\mid\rightarrow 0^+$.\\
Summarising, an inflation period in $N\rightarrow -\infty$ implies $H^2<<\rho_m<<\mid\rho_r\mid$. In these conditions, we finally examine the possible behaviours of $\rho_r$.
\begin{itemize}
\item $\rho_r$ cannot be such that $\rho_r\rightarrow 0^-$ since in $N\rightarrow -\infty$, $\rho_m<<\mid\rho_r\mid$ and $\rho_m$ diverges.
\item If $\rho_r\not\rightarrow 0$ in $N\rightarrow -\infty$, then, from equations (\ref{rhord}-\ref{rhodd}) we get that $\mid \rho_d' \mid<<\mid \rho_r' \mid$ since $\rho_r<0$. It follows that in $N\rightarrow -\infty$, $\mid \rho_d \mid<<\mid \rho_r \mid$. Since we also have $\rho_m<<\mid \rho_r\mid$, then $H^2\simeq \rho_r<0$ in $N\rightarrow -\infty$, that is impossible and also contradicts the above inequality $H^2<<\mid \rho_r\mid$.
\end{itemize}
\end{itemize}
\end{itemize}
Hence, the above behaviours of $y_1$, $\rho_m/\rho_r$ and $\rho_r$ in $N\rightarrow -\infty$ disagree with inflation at that time.\\
As a second step, we now consider that $y_1$ and $\rho_m/\rho_r$ do not tend to a constant or diverge in $N\rightarrow -\infty$. These quantities are thus oscillating and such that their derivatives have an infinity of zero. Let us show that this is not compatible with inflation in $N\rightarrow -\infty$.\\
Equation (\ref{eq1}) indicates that this oscillating behaviour is possible for $y_1$ only when
\begin{itemize}
\item $y_1=1/3(3-4y_2)$. But since $y_2\rightarrow -\infty$ during inflation, it would mean that $y_1$ will diverge, contradicting the above assumption.
\item $y_1=0$. But since $y_1\geq 0$, it would mean that $y_1$ should oscillate with an infinity of minima in $y_1=0$. However, $y_1\propto\rho_m/H^2\propto e^{-3N}/H^2$ and it would imply that $H$ should be discontinuous, diverging each time $y_1=0$, that is an unphysical behaviour for $H$.
\end{itemize}
Hence $y_1$ cannot oscillate and stay finite during an inflation epoch in $N\rightarrow -\infty$. Concerning $\rho_m/\rho_r=y_1/y_2$, an oscillating behaviour is possible only if $(\rho_m/\rho_r)'$ or equivalently, $y_1'y_2-y_1y_2'=y_1(y_2-q)$ has an infinity of zero in $N\rightarrow -\infty$. It will be the case each time
\begin{itemize}
\item $y_2=q$. But $y_2<0$ during an inflation period whereas we assume $q>0$. This case is thus excluded by this last assumption.
\item $y_1=0$. And once again, it would imply an unphysical behaviour for $H$ that should be discontinuous.
\end{itemize}
Hence $\rho_m/\rho_r$ cannot oscillate during an inflation epoch in $N\rightarrow -\infty$.\\
It follows that the possible behaviours of $y_1$, $\rho_m/\rho_r$ or/and $\rho_r$ in $N\rightarrow -\infty$ disagree with an inflation epoch at that time. 
\section{Finite scale factor singularity when dark energy is coupled to matter} \label{a2}
We want to show that an inflation epoch cannot take place in $N\rightarrow -\infty$ but for a special case when $\rho_m/\rho_r$ tends to $-4/3$. We first consider the behaviours of $y_2$ and $\rho_r/\rho_m$ when these two quantities asymptotically diverge or tend to a constant in $N\rightarrow -\infty$. We then show that no behaviour of $\rho_m$ is in agreement with an inflation period taking place in $N\rightarrow -\infty$ but when matter density behaves like radiation density with the opposite sign. In a second time, we assume that $y_2$ and $\rho_r/\rho_m$ do not tend to some constants or diverge in $N\rightarrow -\infty$ and are thus oscillating. Then, we show that these oscillations are not in agreement with an inflation epoch in $N\rightarrow -\infty$.\\ 
Let us begin by considering that $y_2$ and $\rho_r/\rho_m$ asymptotically diverge or tend to some constants in $N\rightarrow -\infty$.
\begin{itemize}
\item $y_2$ cannot tend to a non vanishing constant when $N\rightarrow -\infty$ because, from (\ref{y2}), it follows that $H^2\rightarrow \rho_r$ and there is thus no inflation.
\item $y_2$ cannot tend to a vanishing constant when $N\rightarrow -\infty$ because, since we assume an early time inflation epoch with $y_1<0$, it would mean that $y_2'<0$. But we cannot have $y_2\rightarrow 0$ with $y_2'<0$ in $N\rightarrow -\infty$ since then $y_2$ would be negative, approaching zero from below.
\item It follows from the two previous points that the only behaviour of $y_2$ in agreement with an inflation period in $N\rightarrow -\infty$ is $y_2\rightarrow +\infty$ and thus $\rho_r>>H^2$. We then examine the possible behaviour of $\rho_r$ with respect to $\rho_m$.
\begin{itemize}
\item Let us assume that $\rho_r/\rho_m\rightarrow \alpha+\epsilon_0(N)$ in\footnote{We need to consider a small perturbation $\epsilon_0(N)$ because its integral intervenes in the special case $\alpha =-3/4$.} $N\rightarrow -\infty$ with $\alpha$ a non vanishing and negative (since $\rho_m$ and $\rho_r$ have opposite signs) constant and $\epsilon_0<<\alpha$. Then $y_2\rightarrow \alpha y_1+\epsilon(N)$ with $\epsilon(N)<<y_1$. When $\alpha\not =-3/4$ and since $y_1$ diverges during inflation, it comes from (\ref{eq5}) that $1/y_2\simeq - \frac{3+4\alpha}{\alpha}N$. But then, when $N\rightarrow -\infty$, $y_2$ and thus $y_1$ tend to vanish that disagrees with inflation. If now $\alpha =-3/4$, from (\ref{eq5}) we deduce that $y_2\propto e^{-4N+4\int\epsilon dN}>0$ and thus $y_1\propto-4/3e^{-4N+4\int\epsilon dN}<0$. Introducing these quantities in (\ref{eq4}), one shows that $q\propto e^{-4N+4\int\epsilon dN}>0$ that agrees with our assumption on this quantity. This last expression also means that $y_1\propto -q$ in $N\rightarrow -\infty$ and thus $H\rho_m\propto Q$. It follows from equation (\ref{rhomd1}) that $\dot\rho_m\propto Q$. Then, comparing with equation (\ref{rhodd1}), it comes that when $N\rightarrow -\infty$, $\dot\rho_m\propto \dot\rho_d$ and we may have at early time during inflation $\rho_d\propto\rho_m\propto\rho_r$. This is an interesting possibility. One could conclude that $H^2\propto\rho_r$ and there is no inflation but this is not necessarily the case. Indeed the dominant terms of each $\rho_i$ in the sum $\rho_m+\rho_r+\rho_d$ have not all the same signs. They can thus cancel each other in $N\rightarrow -\infty$, leaving some smaller terms that allow an accelerated expansion at early time. A simple example is given by the asymptotical form of the Hubble function in $N\rightarrow -\infty$, $H=-1/N+\Lambda$ with $\Lambda$ a positive constant. Then in $N\rightarrow -\infty$, Universe expansion accelerates whereas asymptotically all the densities diverge: $\rho_d>0\propto\rho_m<0\propto\rho_r>0\propto e^{-4N}$ with also (since $H$ tends to a constant) $Q\propto e^{-4N}>0$ . Hence, in this special case, inflation can take place at early time in $N\rightarrow -\infty$.
\item $\rho_r/\mid \rho_m\mid$ cannot be such that $\rho_r/\mid \rho_m\mid\rightarrow +\infty$. Then $y_2>>\mid y_1\mid$ and since $y_2\rightarrow +\infty$, (\ref{eq3}) implies that there is no inflation.
\item From the two previous points, the last possibility that we have to consider for $\rho_r$ in $N\rightarrow -\infty$ is thus $\rho_r/\mid \rho_m\mid\rightarrow 0^+$. Hence inflation could also occur in $N\rightarrow -\infty$ when $H^2<<\rho_r<<\mid\rho_m\mid$. In these conditions, the possible behaviours of $\rho_m$ are
\begin{itemize}
\item $\rho_m$ cannot be such that $\rho_m\rightarrow 0$ in $N\rightarrow -\infty$ since we have that $\rho_r<<\mid\rho_m\mid$ but $\rho_r$ diverges.
\item If $\rho_m\not\rightarrow 0$ in $N\rightarrow -\infty$, then from equations (\ref{rhomd1}) and (\ref{rhodd1}) it follows that $\mid \rho_d' \mid<\mid \rho_m' \mid$ since $\rho_m<0$. Hence in $N\rightarrow -\infty$, $\mid \rho_d \mid<<\mid \rho_m \mid$. But we also have $\rho_r<<\mid\rho_m\mid$ and consequently $H^2\simeq \rho_m<0$ that is impossible and also contradicts the above assumption $H^2<<\mid\rho_m \mid$.
\end{itemize}
\end{itemize}
\end{itemize}
Hence, the above behaviours of $y_2$, $\rho_r/\rho_m$ and $\rho_m$ are in agreement with inflation in $N\rightarrow -\infty$ only when $\rho_m\rightarrow -4/3\rho_r<0$. Let us now consider that $y_2$ and $\rho_r/\rho_m$ are some oscillating functions when $N\rightarrow -\infty$, that do not tend to some constants or diverge. Their derivatives have thus an infinity of zero. For $y_2$ it follows from (\ref{eq5}) that this is only possible when 
\begin{itemize}
\item $y_2=1/4(4-3y_1)$. But since $y_1\rightarrow -\infty$ during inflation, it would mean that $y_2$ will diverge and will thus not stay finite when oscillating.
\item $y_2=0$. Since $y_2>0$, it means that $y_2$ should oscillate with an infinity of minima in $y_2=0$. But it seems physically impossible since $y_2\propto\rho_r/H^2\propto e^{-4N}/H^2$ and $H$ should thus be discontinuous, diverging each time $y_2=0$.
\end{itemize}
Hence $y_2$ cannot oscillate and stay finite during an inflation epoch in $N\rightarrow -\infty$. The derivative of $\rho_r/\rho_m=y_2/y_1$ will have an infinity of zero if $y_1y_2'-y_1'y_2=-y_2(y_1+q)$ also has an infinity of zero in $N\rightarrow -\infty$. It could be the case if
\begin{itemize}
\item $y_1=-q$. However, in this case (\ref{eq4}) and (\ref{eq5}) show that $y_1'=y_2'$ and thus $y_1y_2'-y_1'y_2=y_1'(y_1-y_2)$. Since $y_1<0$ and $y_2>0$, this last expression vanishes only when $y_1'=0$. It implies that the infinity of points $(y_1,y_2)$ that makes the derivative of $\rho_r/\rho_m$ vanishing during an inflation epoch, thus allowing this quantity to oscillate, should correspond to equilibrium points of type $P$ as defined in section \ref{s22}. But these kind of points do not allow an accelerated expansion. Hence, an oscillating behaviour of $\rho_r/\rho_m$ in $N\rightarrow -\infty$ when $y_1=-q$ disagrees with inflation.
\item $y_2=0$. But, as explained above, it would imply an unphysical behaviour for $H$ that should be discontinuous.
\end{itemize}
Hence $\rho_r/\rho_m$ cannot oscillate and stays finite during an inflation epoch in $N\rightarrow -\infty$.\\
Consequently, the possible behaviours of $y_2$, $\rho_r/\rho_m$ or/and $\rho_m$ in $N\rightarrow -\infty$ disagree with an inflation epoch at that time but when $\rho_m\rightarrow -4/3\rho_r$.
\bibliographystyle{unsrt}

\begin{thebibliography}{10}
\bibitem{Rie98}
A. G. Riess et al.  [Supernova Search Team Collaboration],
\newblock Astron. J. 116, 1009 (1998).

\bibitem{Per99}
S. Perlmutter et al.  [Supernova Cosmology Project Collaboration],
\newblock Astrophys. J.  517, 565 (1999).

\bibitem{Eis05}
D. J. Eisenstein et al.,
\newblock Astrophys. J. 633, 560 (2005). 

\bibitem{Ade13A}
P. A. R. Ade et al [Planck Collaboration],
\newblock arXiv:1303.5076 (2013).

\bibitem{Gut81}
A. Guth,
\newblock Phys. Rev. D23, 2, 347 (1981). 

\bibitem{Fre87}
K. Freeze et al,
\newblock Nucl.Phys. B287 (1987). 


\bibitem{Wei89}
S. Weinberg,
\newblock Rev. Mod. Phys., 61, 1 (1989). 


\bibitem{Car01}
S. M. Carroll,
\newblock Living Rev. Relativity, 4, (2001).

\bibitem{Sol13}
J. Sola,
\newblock arXiv:1306.1527 [gr-qc] (2013). 


\bibitem{Bra01}
R. H. Brandenberger,
\newblock proceedings BROWN-HET-1256 (2001). 

\bibitem{Lin04}
A. Linde,
\newblock  	Phys.Scripta T117:40-48 (2005). 

\bibitem{Bra89}
R. H. Brandenberger and C. Vafa,
\newblock Nucl. Phys. B 316, 391 (1989). 

\bibitem{Pet08}
P. Peter, N. Pinto-Neto,
\newblock Phys.Rev.D78:063506 (2008). 

\bibitem{Pou13}
A. Pourtsidou et al,
\newblock  	Phys. Rev. D88:083505 (2013). 

\bibitem{Pet13}
V. Pettorino,
\newblock  	Phys. Rev. D88:063519 (2013). 

\bibitem{Ell89}
J. Ellis et al,
\newblock  Phys. Lett. B 228, 264 (1989). 

\bibitem{Ame99}
L. Amendola,
\newblock Phys. Rev. D62:043511 (2000). 

\bibitem{Zim01}
W. Zimdahl et al,
\newblock Phys. Lett. B521:133-138 (2001).

\bibitem{Ame01}
L. Amendola et al.,
\newblock Phys. Rev. D64:043509 (2001).

\bibitem{Ame06}
L. Amendola et al.,
\newblock Phys. Rev. D75, 083506 (2007). 

\bibitem{Muk05}
V. Mukhanov,
\newblock Physical foundations of cosmology, Cambridge University Press (2005).

\bibitem{Bas09}
M. Bastero-Gil and A. Berera,
\newblock Int.J.Mod.Phys.A24:2207-2240 (2009).

\bibitem{Fan80}
L.Z. Fang,
\newblock Phys. Let. B:95, 1, 154 (1980).

\bibitem{Mos85}
I. G. Moss,
\newblock Phys. Lett. B 154, 120 (1985).

\bibitem{Yok88}
J. Yokoyama and K. I. Maeda,
\newblock Phys. Lett. B 207, 31 (1988).

\bibitem{Ber06}
A. Berera,
\newblock  	Contemporary Physics 47, 33 (2006).

\bibitem{BilCol00}
A. P. Billyard and A. A. Coley,
\newblock Phys. Rev. D 61, 083503 (2000).

\bibitem{Pop11}
N. J. Poplawski,
\newblock Phys. Lett. B694, 181-185 (2010).

\bibitem{Ade13}
P. A. R. Ade et al [Planck Collaboration],
\newblock arXiv:1303.5082 (2013).

\bibitem{Lim12}
J. A. S. Lima et al.,
\newblock Phys. Rev.D86:103534 (2012)

\bibitem{Lim13}
J. A. S. Lima et al.,
\newblock To be published in MNRAS (2013)

\bibitem{Gar00}
J. Garriga et al.,
\newblock Phys.Rev.D61:023503 (2000)

\bibitem{Kuo97}
C.-I Kuo,
\newblock Nuovo Cim. B112, 629-632 (1997)

\bibitem{For97}
L. H. Ford et T. A. Roman,
\newblock Phys. Rev. D 55, 2082-2089 (1997)

\bibitem{Pfe98}
M. J. Pfenning et L. H. Ford,
\newblock Doctoral Disseration, arXiv:gr-qc/9805037 (1998)

\bibitem{Lid06}
A. R. Liddle, L. A. Ureña-López,
\newblock Phys.Rev.Lett.97:161301 (2006)

\bibitem{Arb09}
A. Arbey,
\newblock proceedings of the conference "Invisible Universe", Paris, June 29 - July 3, France, (2009)

\bibitem{Mor13}
S. C. F. Morris et al.,
\newblock arXiv:1304.2196 (2013)

\bibitem{Hry10}
O. Hrycyna and M. Szydlowski,
\newblock JCAP 1012:016 (2010)

\bibitem{Wan13}
C. H.-T. Wang et al.,
\newblock arXiv:1309.4066

\bibitem{Pav09}
D. Pavón and B. Wang 
\newblock Gen. Rel. Grav., 41:1 (2009)

\bibitem{Wri07}
E. L. Wright,
\newblock A.P.J., 664:633-639 (2007)
\end{thebibliography}

\end{document}